\documentclass[nofootinbib,twocolumn]{revtex4}
\usepackage{graphicx}
\usepackage{amssymb}
\usepackage{amsmath}
\usepackage{bm}
\usepackage{hyperref}

\begin{document}

\title{Aspects of the cosmological ``coincidence problem"}

\author{H.E.S. Velten}\email{velten@pq.cnpq.br}\author{R.F. vom Marttens}\email{rodrigovommarttens@gmail.com}\author{W. Zimdahl}\email{winfried.zimdahl@pq.cnpq.br}

\affiliation{Departamento de F\'isica, Universidade Federal do Esp\'{\i}rito Santo (UFES), Av. Fernando Ferrari, 514, Campus de Goiabeiras, CEP 29075-910,
Vit\'oria, Esp\'{\i}rito Santo, Brazil}

\begin{abstract}
The observational fact that the present values of the densities of dark energy and dark matter are of the same order of magnitude, $\rho_{de0}/\rho_{dm0} \sim \mathcal{O}(1)$, seems to indicate that we are currently living in a very special period of the cosmic history. Within the standard model, a density ratio of the order of one just at the present epoch can be seen as coincidental since it requires very special initial conditions in the early Universe.
The corresponding ``why now" question constitutes the cosmological ``coincidence problem". According to the standard model the equality $\rho_{de} = \rho_{dm}$ took place ``recently'' at a redshift $z \approx 0.55$.
The meaning of ``recently'' is, however, parameter dependent. In terms of the cosmic time the situation looks different. We discuss several aspects of the ``coincidence problem", also in its relation to the cosmological constant problem, to issues of structure formation and to cosmic age considerations.
\\


\end{abstract}

\maketitle

\section{Introduction}

To study cosmology, the most common approach is to adopt general relativity equipped with the homogeneous, isotropic and expanding Friedmann-Lema\^{\i}tre-Robertson-Walker (FLRW) metric. This results in the Friedmann equations which relate the dynamics of the Universe to its matter-energy content. Therefore, it is clear for any astronomer that  observations of the cosmic dynamics allow us to infer the Universe's composition.

Modern astronomy led us to the so called standard cosmological model in which $\sim$ 70\% of the today's cosmic energy budget (roughly the critical energy density of the Universe $\rho_{c}\sim 10^{-29}$ g/cm$^3$) correspond to an apparently mysterious component called dark energy. In its  most accepted form it is identified with a cosmological constant $\Lambda$. The remaining contributions are composed of matter (dark matter and baryonic matter),  accounting for $\sim$ 30\%, and an almost negligible radiation (photons) contribution of the order of $\sim 10^{-3}$\%. This configuration is summarized by the acronym $\Lambda$CDM  where CDM stands for cold dark matter. However, the energy distribution on the components has been changing through the Universe's history.

Due to the fact that the densities of the components scale in different ways, the cosmic history can be divided into three different epochs (not taking into account here an early inflationary phase). According to the hot big-bang model, the ``initial state" of the Universe was that of a dense and hot expanding fireball. This means that the dynamics of the Universe at that stage was determined by the radiation component which dominated the total energy content. But as the Universe expands and cools, the radiation density drops faster than the dark-matter density. At a redshift $z_{eq}\sim 3400$ the equality of radiation and matter densities occurs. This is the beginning of the matter-dominated epoch and, after this moment, dark matter drives the expansion. During most of its life, the Universe made use of this matter dominance to form structures like stars, galaxies and galaxy clusters by gravitational instability. This process would be endless if a third epoch had not arrived ``recently". This is the current dark-energy phase which started at the moment where the matter density had dropped to the same value as the dark-energy density. According to the current view, this happened at a redshift $z_{de}\sim 0.55$. Since then the Universe experiences an accelerated expansion phase where gravity is no longer able to efficiently form super-galaxy clusters.

The nature of both dark energy and dark matter is still unknown. The prevailing opinion assumes dark energy to be a cosmological constant $\Lambda$ whilst dark matter is modeled as a nonrelativistic fluid.

The remarkable fact that the energy densities of dark energy and dark matter are of the same order around the present time seems to indicate that we are living in a very special moment of the cosmic history. Within the standard model where the dark-energy density is constant and the dark-matter density scales with the inverse third power of the cosmic scale factor this appears to be a coincidence since it requires extremely fine-tuned initial conditions in the early Universe. Both in the very early Universe and in the far-future Universe these energy densities differ by many orders of magnitude.
This so called ``Cosmological Coincidence Problem'', hereafter CCP, was first formulated in Steinhardt's contribution to the proceedings of a conference celebrating the 250th anniversary of Princeton University \cite{steinhardt}. Since then, ``the coincidence problem'' became a common jargon in the cosmological literature. Many textbooks and review papers have addressed the CCP.
As an example we quote from D'Inverno's classical general relativity book, ``{\it whenever we find coincidences in a physical theory, we should be highly suspicious about the theory.}'' \cite{Dinverno}.

One may also formulate the CCP in terms of the ratio of the energy density of dark matter to the dark-energy density. This ratio changes from almost infinity to zero in the $\Lambda$CDM model. Our present period then appears to be singled out by a ratio of the order of one.

The CCP has motivated the creation of many different dark-energy models which pretend to ``explain" why, only ``now'', the matter density, after it has dropped many orders of magnitude, has reached exactly the same value as the dark-energy density. In general, ``solutions'' to the CCP involve a non-standard behavior of scalar-field type dark energy \cite{DECCP} or interaction models in which dark-matter and dark-energy densities are of the same order for a significant fraction of the lifetime of the Universe \cite{IntCCP}. A different line of thinking relies on anthropic considerations in which conditions for the existence of observers (us) in an ensemble of astronomers set upper bounds on the dark-energy density \cite{weinberg87, AnCCP}.

On the other hand, one may wonder, why such coincidence is seen as a big problem. There are indeed many scientists in the field who deny this \cite{rovelli}.
It appears to be a problem if one starts assuming (tacitly) that we could find ourselves with equal probability in any of the periods of the cosmic evolution. But this is definitely not the case. Interestingly \cite{shawbarrow}, the time scale defined by the (effective) cosmological constant is not only of the order of the present age of the Universe but also of the same order as the scale that is relevant for heavier elements being produced.
So, anthropic arguments necessarily enter the discussion.

But even if one disregards this obvious fact, the severity of the problem will depend on whether we choose a linear or a logarithmic time scale as will be visualized in Figs.~1--4.

Namely, on a linear time scale, it is during a considerable time span of the cosmic evolution up to now that the ratio of the energy densities is not so much different from unity.
The problem would not be that big either if the community could accept a density ratio of the order of one  as a consequence of using Friedmann's equations with a non-vanishing cosmological constant $\Lambda$ as a natural ingredient of Einstein's theory \cite{rovelli}. This $\Lambda$ would play the role of another gravitational constant along with Newton's gravitational constant $G$. And, in a classical (non-quantum) theory, this $\Lambda$ has the value that it has, in the same sense that $G$ has the value that it has. But this does not seem to be the generally accepted view. The reason is that, based on quantum field theoretical considerations in Minkowski space, the quantum vacuum contributes to the energy-momentum tensor on the right-hand side of Einstein's equations in a formally similar way as the cosmological constant contributes on the left-hand side of the field equations. Then, the observed cosmological constant is not just the original purely geometrical $\Lambda$ but the combination of this quantity with the quantum vacuum part and the CCP is related to the resulting effective cosmological term.

Alternatively, setting (without further motivation) Einstein's cosmological constant to zero, i.e. $\Lambda = 0$, attributes the observed accelerated expansion of the Universe entirely to quantum vacuum effects. Although it is an open question whether or how the quantum vacuum acts gravitationally, there has been a lot of research devoted to phenomenological models in which the vacuum energy density is no longer constant but changes in time which implies a non-gravitational interaction with dark matter (see, e.g., \cite{saulo,sola} and references therein).

The simplest constraint on the (effective) cosmological ``constant" is that it should not prevent the formation
of galaxies \cite{weinberg87}. But it should also guarantee an age of the Universe that is higher than the age of
globular clusters which are considered to be the oldest observed objects in our Universe \cite{krauss, globular}.
The latter requirement is not consistent with the traditional Einstein-de Sitter universe in which there is no cosmological constant.

As already mentioned, a coupling between dark matter and dark energy is frequently introduced to address the CCP. The general aim here is to find a mechanism which dynamically explains a fixed ratio of the densities of dark matter and dark energy. Examples are tracker \cite{tracker}, attractor \cite{attractor} or other fixed point solutions \cite{dodelson,lip,annap}.
There are solutions in which dark energy dominates periodically and therefore it is nothing special if we find
ourselves in an accelerating part of the cycle \cite{lip,annap}.

If the Universe really approaches a stationary or periodic solution with finite values of the density parameters close to the presently observed values, this would certainly solve the CCP.

While it is not impossible to construct solutions of these types, the interactions that are required to produce them seem artificial so far since they lack a real physical motivation.
The CCP is not really ``solved" this way as sometimes claimed but the original aim to explain a ratio of the order of one of the energy densities is replaced by the necessity to explain a phenomenologically introduced coupling in the dark sector with an appropriate strength to fit the observational data.
Anyway, these attempts have their merits as models competing with the standard model even without the motivation to ``solve" or to soften the CCP.

It has also been argued that the coincidence problem is ameliorated in a phantom-dominated universe.
Because of its finite lifetime the densities of the dark components are comparable for a larger fraction of the total lifetime of the Universe \cite{scherrer}.

The goal of this paper is to put together some ideas and arguments concerning the CCP and to illustrate several of its aspects by simple graphic representations. This paper has the following structure: in the next section, Sec.~\ref{exp}, we recall basics of the homogeneous and isotropic background expansion. This serves as the presentation of the CCP itself. Then, in Sec.~\ref{texp}, we argue that one of the issues behind the CCP depends on the choice of the redshift $z$ as our temporal parameter. If we use instead a parametrization in terms of the cosmic time the coincidence appears much less dramatic. This is indeed a main argument by many cosmologists who do not consider the CCP to be a central problem in cosmology.
However, the obvious possibility of a different time parametrization is only one aspect. The value of the
energy-density ratio is also crucial for structure formation and for the age of the Universe as we shall discuss
in Sec.~\ref{structure}.
A brief summary is given in section \ref{final}.

\section{The expansion of the universe}
\label{exp}

Edwin Hubble's work in the late 1920s was crucial for the development of modern cosmology. Firstly, his observations were important to certify the existence of galaxies other than our own galaxy, the Milky Way. Then, by increasing the number of observed galaxies, he established a key ingredient for any modern cosmology which is the expansion of the Universe. Hubble realized that galaxies move away from each other with a velocity proportional the their distance. To see this one has to parameterize
the expansion of the Universe in terms of the cosmic scale factor $a(t)$. Therefore, the physical distance ${\bf r_{phys}}$ between two objects at a certain time $t$ can be written as ${\bf r_{phys}}= a(t) {\bf r_0}$, where ${\bf r_0}$ is a fixed distance at some time $t_0$ which we identify with the present time.
Here and throughout, a subindex $0$ denotes the present value of the corresponding quantity.  Calculating ${\bf \dot{r}_{phys}}$ gives
\begin{equation}
\label{Hubble}
\mathbf{\dot{r}_{phys}} \equiv {\bf v_{phys}}=H(t) {\bf r_{phys}},
\end{equation}
where $H$ is the Hubble rate
\begin{equation}\label{Hat}
H(t)=\frac{\dot{a}(t)}{a(t)}.
\end{equation}
Taken at the present time with $H(t_{0}) \equiv H_{0}$, relation (\ref{Hubble})
is the famous Hubble law, $H_{0}$ being the Hubble constant.
In fact, although this law bears Hubble's name it already appears in an earlier paper by Lema\^{\i}tre \cite{lemaitre}.

It is useful to normalize the value of the scale factor today to  unity, i.e., $a_0=1$.  With this convention the scale factor is related to the redshift parameter $z$ via $a(t)=(1+z)^{-1}$.

General relativity is the standard theory of the gravitational field, embodied by
Einstein's equations
\begin{equation}\label{Eeq}
R_{\mu\nu}-\frac{1}{2}g_{\mu\nu} R + \Lambda g_{\mu\nu}=8\pi G T_{\mu\nu},
\end{equation}
where $g_{\mu\nu}$ is the metric tensor, $R_{\mu\nu}$ is the Ricci tensor and $R$ is the Ricci scalar. $\Lambda$ is the cosmological constant. The entire left-hand side of Eq.~(\ref{Eeq}) represents the space-time geometry.
The matter part is taken into account by the energy-momentum tensor $T_{\mu\nu}$ on the right-hand side. $G$ is Newton's gravitational constant.
Einstein's equations quite generally describe the interplay between the space-time geometry and the matter distribution.
A successful method to find solutions of these equations is to impose a symmetry adapted to the problem at hand.
In standard cosmology such a symmetry is provided by the cosmological principle according to which our Universe
at the largest scales is spatially homogeneous and isotropic. These symmetry requirements fix the metric to be of the structure of the FLRW metric
\begin{equation}\label{flrw}
ds^2= -dt^2+a(t)^2\left[\frac{dr^2}{1-kr^{2}}+d\Omega^2\right],
\end{equation}
where the constant $k$ represents the spatial curvature.
As usual, we use units in which the velocity of light is unity.
Einstein's equations then determine the scale factor $a(t)$.
Use of (\ref{flrw}) in (\ref{Eeq}) leads to the Friedmann equation
\begin{equation}\label{Fried}
H^2-\frac{k}{a^2}=\frac{8\pi G}{3} \sum_i \rho_i + \frac{\Lambda}{3},
\end{equation}
where $\rho_i$ is the energy density of the $i$th matter component. For each of these components one may define a fractional density parameter
\begin{equation}
\Omega_{i}(z) =\frac{\rho_i(z)}{\rho_{c0}}, \hspace{0.4cm} {\rm with} \hspace{0.4cm} \rho_{c0}=\frac{3}{8\pi G} H^2_0,
\end{equation}
where $\rho_{c0}$ is the today's critical density. From this definition we also have the today's value of the fractional density of component $i$, $\Omega_{i0}=\rho_i(0)/\rho_{c0}$.
In a similar way one defines the fractional quantities
\begin{equation}\label{}
\Omega_{k0} = -\frac{k}{H_{0}^{2}} \hspace{0.4cm} {\rm and } \hspace{0.4cm}
\Omega_{\Lambda 0} = \frac{\Lambda}{3H_{0}^{2}}.
\end{equation}
Assuming the material content of the Universe to consist of radiation (subindex $i=r$) and matter (subindex $i=m$),
Eq.~(\ref{Fried}) can be rewritten as
\begin{eqnarray}
H^2(z)&=&H^2_0\left[\Omega_{r0}(1+z)^4 +\Omega_{m0}(1+z)^3 \right.\nonumber\\
&& \left. \qquad\qquad+\Omega_{k0}(1+z)^2 + \Omega_{\Lambda 0}\right].
\end{eqnarray}
From this expression a reference model can be constructed by using the parameter values reported by the Planck team ({\it Planck + WMAP}) which are $\Omega_{m0}=0.315$,  $\Omega_{\Lambda 0}=0.685$ and $H_0=67.3$ km s$^{-1}$ Mpc$^{-1} =6.9832$ x $10^{-26}$ year$^{-1}$. Note that in the $\Lambda$CDM scenario $\Omega_{m0}$ means the sum of both dark matter $\Omega_{dm0}$ and baryonic matter $\Omega_{b0}$ \footnote{We will make a distinction between them only in section \ref{structure}.}. For the curvature one has $\Omega_{k0}=-0.0042^{+0.0043}_{-0.0048}$ which means these data are fully consistent with a spatially flat Universe. The redshift of equality between matter and radiation is $z_{eq}=3391$ which gives $\Omega_{r0}=9.29$ x $10^{-5}$ \cite{planckparameters}.

\begin{figure}\label{fig1}
\begin{center}
\includegraphics[width=0.43\textwidth]{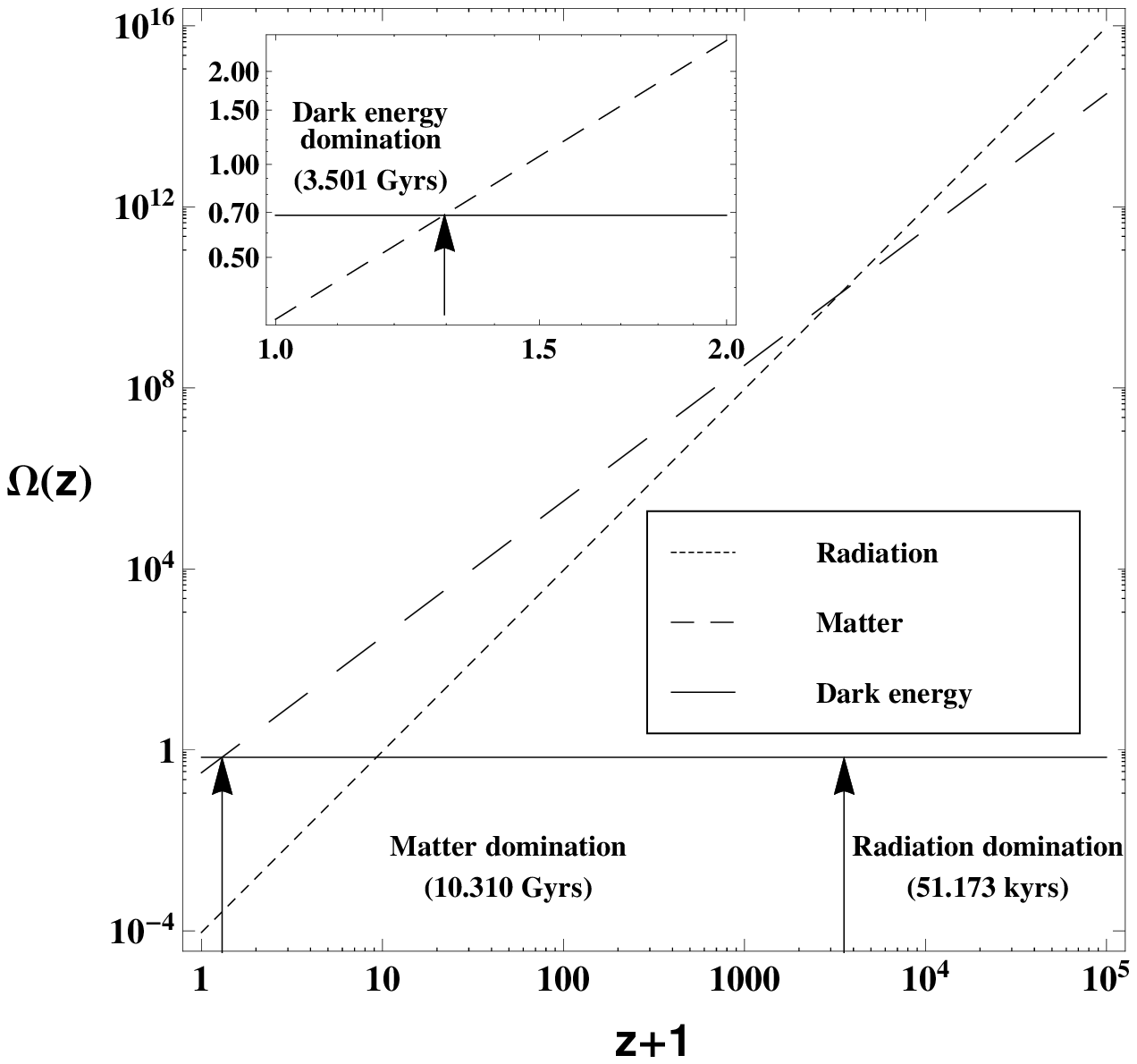}
\caption{Evolution of fractionary energy densities $\Omega$ as a function of redshift.}
\end{center}
\end{figure}

Fig.~1 shows the densities $\Omega_i$ for the components $i=$ matter (long dashed), radiation (short dashed) and dark energy (solid) as a function of the redshift, a plot that is well known in the literature. The matter density becomes of order $\mathcal{O}(1)$ close to the present epoch, but it was $\sim 10^8 \rho_{c0}$ at the time when the Cosmic Microwave Background (CMB) photons were released  at $z_{CMB}\sim 1000$. The curvature fraction has been neglected.

There is another way to express the fractionary parameters. Sometimes they are defined in terms of the time-dependent critical density as
\begin{equation}\label{Omegarhoc}
\Omega^{\star}_{i}(z) =\frac{\rho_i(z)}{\rho_{c}(z)}, \hspace{0.4cm} {\rm with} \hspace{0.4cm} \rho_{c}(z)=\frac{3}{8\pi G} H^2(z)
\end{equation}
as well as
\begin{equation}\label{}
\Omega_{k}^{\star} = -\frac{k}{H^{2}(z)} \hspace{0.4cm} {\rm and } \hspace{0.4cm}
\Omega_{\Lambda}^{\star} = \frac{\Lambda}{3H^{2}(z)}.
\end{equation}
With the help of these definitions the Friedmann equation is written as
\begin{equation}\label{Omegarhoc1}
\Omega_{r}^{\star}(z) +\Omega_{m}^{\star}(z)+\Omega_{k}^{\star}(z)+\Omega_{\Lambda}^{\star}(z)=1.
\end{equation}

Obviously, the cosmological constant contribution changes with the expansion. From (\ref{Omegarhoc}) -- (\ref{Omegarhoc1}) we obtain Fig.~2, another common figure to visualize the CCP. Again, the curvature part has been neglected.

\begin{figure}\label{fig2}
\begin{center}
\includegraphics[width=0.42\textwidth]{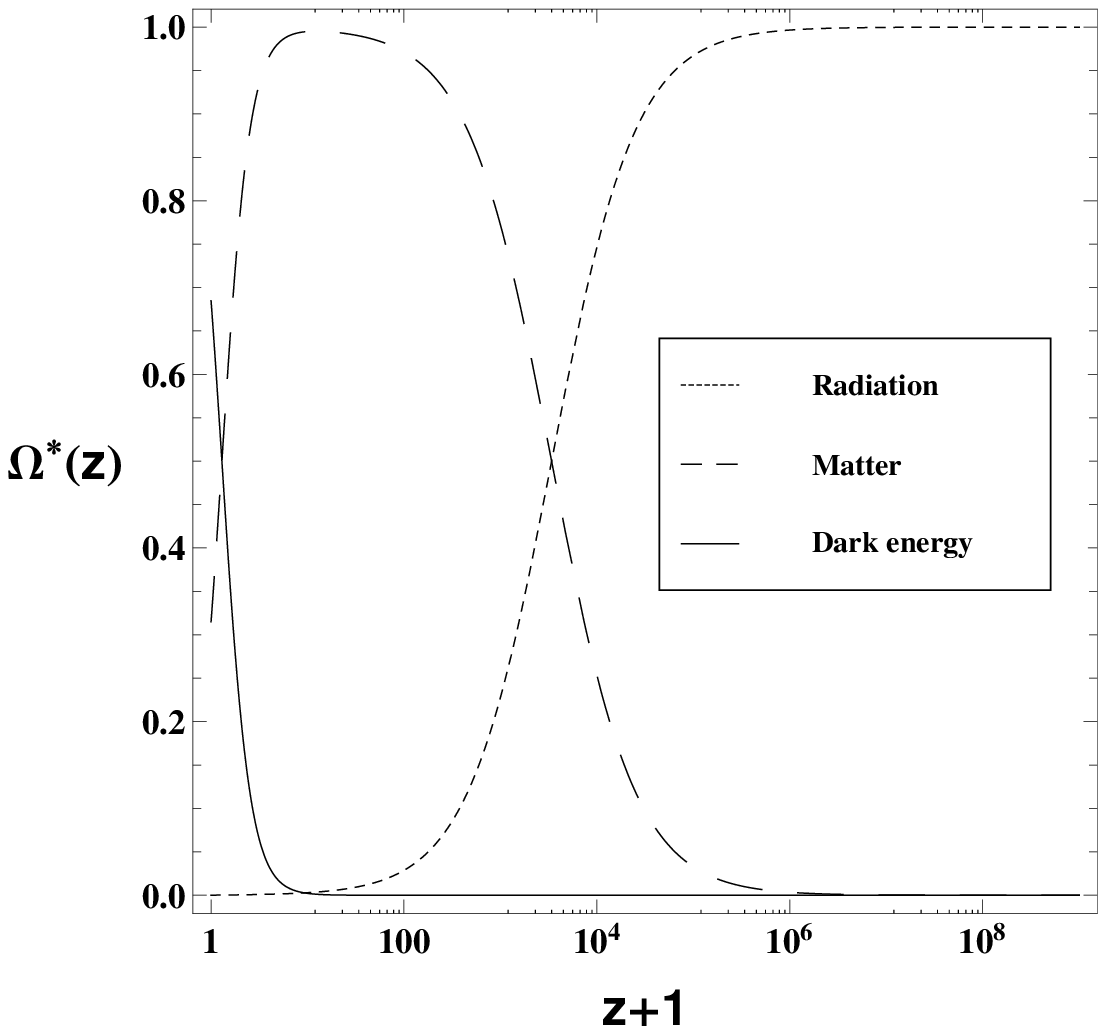}
\caption{Evolution of fractionary energy densities $\Omega^{\star}$ as a function of redshift.}
\end{center}
\end{figure}

From the perspective shown in both Fig.~1 and Fig.~2 the dark-energy dominated epoch occupies only a small fraction of the entire redshift range. These plots suggest indeed that we are living in a very special era of the cosmic
evolution.

\section{The cosmic time $t$ as the expansion variable}
\label{texp}

For historical reasons the redshift $z$ has been used as the most common way to express the distance of cosmic objects.
The redshift is a directly observable quantity that reflects the kinematics of astronomical objects. For the local Universe the relation $v= c z$ (the low-redshift approximation of (\ref{Hubble})) is valid and therefore the redshift of galaxies can be directly translated into velocities and vice versa.

As shown in Fig.~1, the matter energy density  parameter $\Omega_{m}$  crosses the constant $\Omega_{\Lambda}$ value only at a very low $z$ value.  A similar visualization of the CCP in terms of
 $\Omega_{m}^{\star}$ and $\Omega_{\Lambda}^{\star}$ is given in Fig.~2.

Any value of $\Omega_{\Lambda}$ in the interval $0<\Omega_{\Lambda}<1$ (or $0<\Omega_{\Lambda}^{\star}<1$) would give rise to a comparable picture with different crossing redshifts.

Let us now redesign Fig.~1 by using the cosmic time $t$ instead of the redshift $z$. In order to obtain an expression for the cosmic time it is necessary to integrate relation(\ref{Hat}):
\begin{equation}\label{age}
\int^{t}_{0} dt = \int^a_0 \frac{da^{\prime}}{ a^{\prime} H(a^{\prime})}\,.
\end{equation}
With the today's scale factor $a_0=1$, this expression gives an age of the Universe of $13.8$ Gyrs for the standard $\Lambda$CDM model. Changing the limits of integration in (\ref{age}) accordingly, one can also find the look-back time which is the time interval from now to some point in the past as well as the time that has passed between any two events in the Universe. In this way we calculate the duration of the radiation, matter and dark energy eras which are also shown in Fig.~1. The radiation domination is very short in terms of $t$. It lasts only $51.173$ Kyrs ending at a redshift $z_{eq}=3391$. A substantial fraction of the Universe's lifetime is dominated by matter. Indeed, this is a crucial epoch for the development of cosmic structures like stars, galaxies and clusters. This phase can not be much shorter because this would avoid the formation of an environment where life in the Universe can appear.

The dark-energy dominated epoch started $3.5$ Gyrs ago. This is about $1/4$ of the Universe's lifetime. Therefore, having these numbers in mind, the dark-energy epoch did not start so recently as the redshift parametrization suggests.

In Fig.~3 we show the evolution of the fractionary densities using now the cosmic time as our expansion variable. The grey vertical stripe indicates the reionization in the redshift range $15>z>6$ \cite{reion}.
The solar system formation occurred when the density of the dark energy was of similar order than the matter density. Many other astrophysical events happened around this time as, for instance, the onset of the nonlinear regime of large-scale structure formation.
Concerning the latter, the interesting concept of a  backreaction mechanism has been developed. The idea is that the existence of cosmic structures causes a deviation from exact large-scale homogeneity and isotropy, affecting the background cosmic expansion \cite{Back}.
According to this concept, the accelerated expansion might be a consequence of the formation of the first structures. Therefore, in this context there is no CCP since there is a well justified origin for the acceleration.
However, so far there does not seem to exist an agreement on whether or not the backreaction is sufficiently
large to induce an acceleration of the expansion \cite{greenwald}.

\begin{figure}\label{fig3}
\begin{center}
\includegraphics[width=0.45\textwidth]{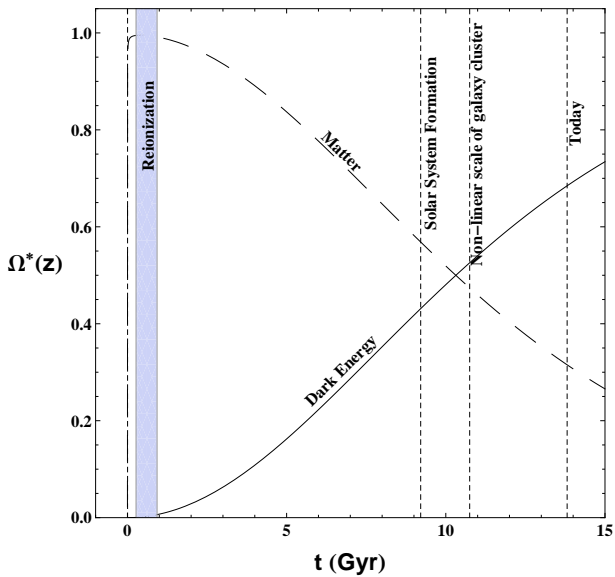}
\caption{Evolution of fractionary energy densities as a function of the cosmic time.}
\end{center}
\end{figure}

Fig.~4 presents a complementary plot in which the variation of the redshift as a function of the cosmic time is shown. For example, for low redshifts, a reference period which lasts $\Delta t =2$ Gyrs correspond to only a small variation in the redshift space of $\Delta z =0.191$. On the other hand, in the range $z \sim 2-3$, the same period of $\Delta t =2$ Gyrs is covered by an interval of $\Delta z = 1.577$. This shows that the relation between intervals of cosmic time and intervals of cosmic redshift changes substantially throughout the expansion of the Universe.

\begin{figure}\label{fig4}
\begin{center}
\includegraphics[width=0.4\textwidth]{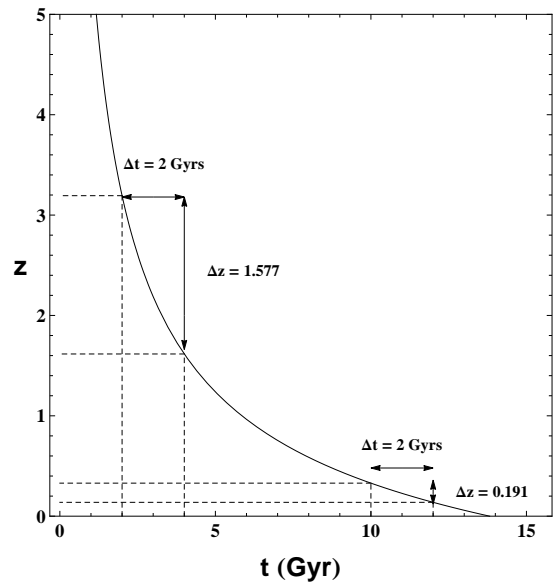}
\caption{Variation of the redshift as a function of the cosmic time.}
\end{center}
\end{figure}

Judged from Figs.~3 and 4  the CCP seems much less pressing than judged from Figs.~1 and 2. In the cosmic-time parametrization the densities of dark energy and dark matter are of a similar order over a substantial fraction of the cosmic history, not just ``recently".
The equality $\Omega_{m}^{\ast} = \Omega_{\Lambda}^{\ast}$ occurs at a redshift of $z=1.157$.
The look-back time at this redshift is $\sim 8.5$ Gyrs which is of the order of half the age of the Universe.

\section{Structure formation and the age of globular clusters}
\label{structure}

The CCP has aspects beyond the question whether or not the equality of the densities of dark energy and dark matter appeared more or less recently. As already mentioned, the ratio of the densities is crucial for structure formation. This can be quantified by asking, how much different the Universe would look like if, instead of $\Omega_{\Lambda}=0.7$, this parameter were $\Omega_{\Lambda}=0.8$ or $\Omega_{\Lambda}=0.6$.
What would be the consequences if one of these values were the ``correct" one?

One of the possible consequences is related to the large scale structure formation process. In this context one usually defines the matter density contrast $\Delta=\hat{\rho}_{m}/\rho_{m}$, where 
$\hat{\rho}_{m}$ is an inhomogeneous density perturbation and $\rho_{m}$ is the homogeneous background density. 
The so called linear regime is characterized by $\hat{\rho}_{m} \ll \rho_{m}$, equivalent to $\Delta \ll 1$. 
The linear density contrast obeys the equation
\begin{equation}\label{deltagrowth}
\ddot{\Delta}+2H\dot{\Delta}+\left(\frac{c^2_s k^2}{a^2}-4\pi G \rho\right)\Delta=0,
\end{equation}
where $c^2_s$ is the square of the speed of sound of the fluid and $k$ is the comoving wavenumber \footnote{This $k$ should not be confused with the previous spatial curvature parameter which was also denoted by $k$. The spatial curvature is assumed to be zero here.}. 
A linear approximation is justified as long the perturbations remain small. For pressureless matter $\Delta$ grows proportional to the scale factor $a$. 
Now, the observed structures in the Universe are strongly nonlinear and an adequate description of their formation cannot be obtained using just equation (\ref{deltagrowth}) which breaks down for $\Delta\approx 1$. 
Nevertheless, equation (\ref{deltagrowth}) can provide us with the desired information. 
Namely, it can tell us whether or not for a certain parameter combination and for suitable initial conditions 
values of the order of $\Delta \approx 1$ can be obtained. 
For parameter values which result in solutions of (\ref{deltagrowth}) for which  $\Delta$ 
remains smaller than one until the present time, structure formation is obviously impossible since there is not sufficient growth to reach the nonlinear regime. 
In other words, the possibility of a perturbation growth until $\Delta\approx 1$ before the present time can be used as a criterion for structure formation.

We can solve equation (\ref{deltagrowth}) with realistic initial conditions for any given scale. Basically, the scale determines the amplitude of the dark-matter density perturbation at the matter-radiation equality, the moment from which perturbations rigorously obey Eq. (\ref{deltagrowth}). As usual, we will assume a flat cosmology with a baryonic contribution of $\Omega_{b0}=0.05$. In the top panel of Fig.~5 we show that for a value $\Omega_{\Lambda}=0.8$ the acceleration stage had started much earlier and, consequently, the largest structures we observe in the Universe like galaxy clusters ($k=0.2 h Mpc^{-1}$) would never have formed, i.e. they would never reach the nonlinear regime $\Delta=1$. Therefore, one could argue that since formation of collapsed structures is a necessary condition for enabling life in the Universe, one can set an upper bound on the value of $\Omega_{\Lambda}$ from anthropic arguments. 

On the other hand, the bottom panel of Fig.~5 shows the corresponding dark-matter perturbation growth for a much smaller scale like proto-galaxies ($k=1000 h Mpc^{-1}$). This is probably one of the smallest scales which can be studied within the realm of cosmology. Then, this seems to be the safest scale on which one can invoke anthropic arguments in order to infer an allowed maximum value for $\Omega_{\Lambda}$. The chosen values $\Omega_{\Lambda}=0.85, 0.9$ and $0.925$ can be interpreted as being cosmologies where the dark-matter density contribution is twice, the same or half of the baryonic matter contribution, respectively. Only the latter scenario ($\Omega_{\Lambda}=0.925$) is definitely inconsistent with the existence of proto-galaxies.

\begin{figure}\label{fig5}
\begin{center}
\includegraphics[width=0.4\textwidth]{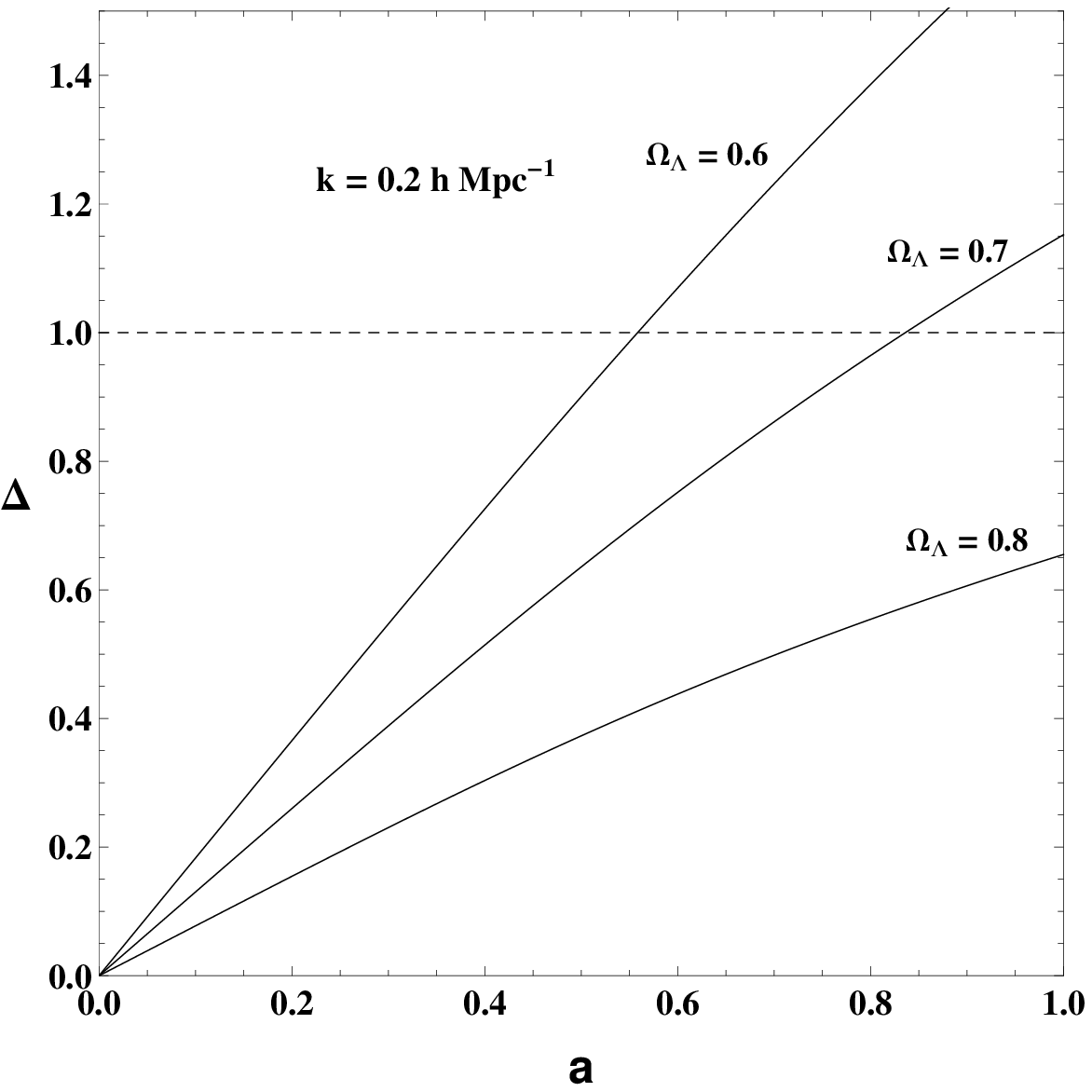}
\includegraphics[width=0.4\textwidth]{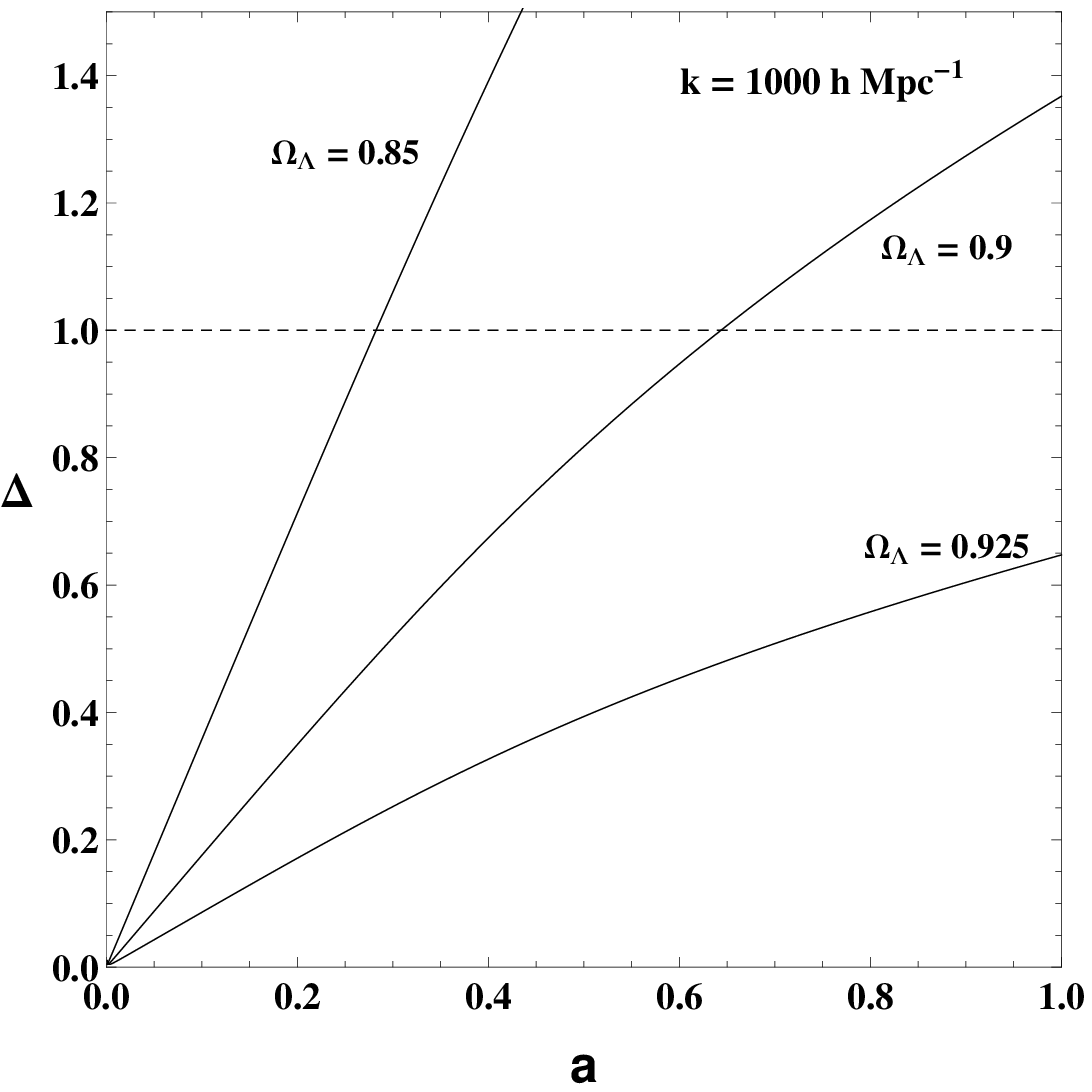}
\caption{Growth of the dark matter perturbations in a $\Lambda$CDM universe as a function of the scale factor $a$. The top panel corresponds to galaxy cluster scales ($k=0.2 h\,Mpc^{-1}$) while the bottom panel corresponds to proto-galaxies ($k=1000 h\,Mpc^{-1}$). Different $\Omega_{\Lambda}$ values have been adopted. The horizontal dashed line sets the onset of the nonlinear regime of structure formation.}
\end{center}
\end{figure}

\begin{figure}\label{fig6}
\begin{center}
\includegraphics[width=0.4\textwidth]{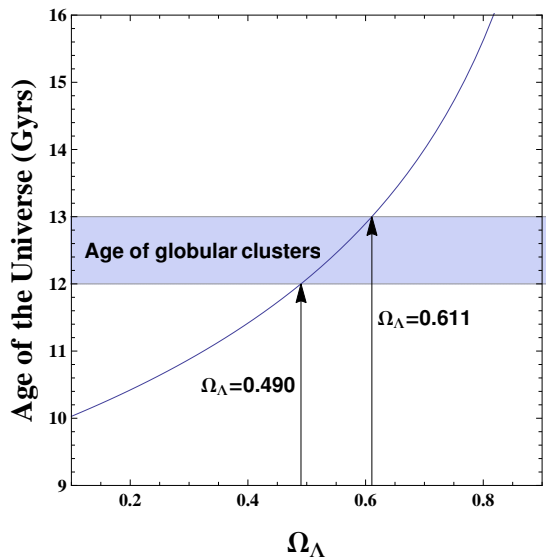}
\caption{Age of the Universe as a function of $\Omega_{\Lambda}$. The horizontal stripe corresponds to the age of some known globular clusters. }
\end{center}
\end{figure}

The case $\Omega_{\Lambda}=0.6$ would apparently not introduce drastic changes compared with $\Omega_{\Lambda}=0.7$. In this situation the matter epoch would last more than $10$ Gyrs. Of course, we would not observe the current magnitude of accelerated expansion and the standard cosmological model would be quantitatively different.
Therefore, from the point of view of structure formation the concordance value $\Omega_{\Lambda}=0.7$
does not seem to be the only possible value to produce a viable cosmology. 
However, here astrophysical considerations come into play. The age of stellar populations like globular clusters plays an important role for setting a lower bound on $\Omega_{\Lambda}$. Current estimations of the oldest objects in the Universe can determine a ``minimum'' allowed age of the Universe. Conservative estimations for some ages reach values of the order of $12-13$ Gyrs \cite{krauss, globular}. Therefore, an Einstein-de-Sitter universe $\Omega_{\Lambda}=0$ corresponding to an age of $\sim 9$ Gyrs is surely ruled out with arguments involving the age of astrophysical objects.
In order to assure that the Universe is at least $12$ Gyrs old one has to assume the value $\Omega_{\Lambda}=0.5$, whereas $\Omega_{\Lambda}=0.6$ leads to $12.7$ Gyrs.
Although the limits set by the age of the oldest astrophysical objects are not very precise, they point to a lower limit of $\Omega_{\Lambda}$ which is close to the standard-model value $\Omega_{\Lambda}=0.7$. Fig.~6 exemplifies this discussion.

Consequently, the joint information from structure formation and from age considerations does indeed restrict $\Omega_{\Lambda}$ to a value very close to the observed $\Omega_{\Lambda} = 0.7$.

\section{Final Discussion}
\label{final}

If the expansion of the Universe is parameterized in terms of the redshift $z$ we seem to live in a very special period of the cosmic history, characterized by a density ratio of dark matter to dark energy of the order of one.
The equality of both densities occurs at a rather low value of $z$ which suggests that the dominance of dark energy started ``recently" or even ``now" on a cosmological scale, giving rise to the CCP problem ``why now?".
This issue became a very popular source of discussion in cosmology and it has motivated the appearance of many cosmological models which try to explain or to alleviate this coincidence.

If, on the other hand, as we have demonstrated in some detail, the cosmic time is used to parameterize the expansion, the equivalence between matter and dark energy densities occurs when the Universe has about $3/4$ of its current age. Since the remaining part of $1/4$ does not represent a very small fraction of the present lifetime of the Universe, there does not seem to be room for a ``why now?" question.

However, this does not mean that the exact present value of the energy-density ratio is of minor importance.
On the contrary. Successful structure formation provides us with an upper limit not much larger than the presently observed $\Omega_{\Lambda} = 0.7$ and the circumstance that the Universe has to be older than the oldest astrophysical objects requires an $\Omega_{\Lambda}$ which cannot be considerably smaller than $\Omega_{\Lambda} = 0.7$. It is this coincidence which deserves a deeper explanation.

\textbf{Acknowledgement}:  We thank Oliver Piattella for important comments on the manuscript. We also thank CNPq and FAPES for financial support.

\end{document}